\documentclass[twocolumn,aps,superscriptaddress,showpacs,nofootinbib,floatfix]{revtex4-1}
\usepackage{graphicx} 
\usepackage{epsfig,bm}
\usepackage{graphicx,amssymb}
\usepackage{amsmath}
\usepackage{algorithm, algpseudocode}
\usepackage{subcaption}
\usepackage{epstopdf}
\usepackage[eps]{pstricks}
\usepackage{slashed}
\usepackage{float}

\usepackage[normalem]{ulem}

\newcounter{subalgorithm}[algorithm]

\renewcommand{\sout}{\bgroup \color{red} \ULdepth=-.5ex \ULset}

\begin{document}

\title{An AI-Inspired Numerical Method in the Quark Model: Application to Finding the Wave Functions for Heavy Tetraquark States}

\author{Daeho Park}
\email{olmu100@yonsei.ac.kr}
\affiliation{Department of Physics and Institute of Physics and Applied Physics, Yonsei University, Seoul 03722, Korea}

\author{Su Houng Lee}\email{suhoung@yonsei.ac.kr}
\affiliation{Department of Physics and Institute of Physics and Applied Physics, Yonsei University, Seoul 03722, Korea}

\date{\today}

\begin{abstract}
The current ongoing advancements in AI have shed light on the landscape of numerical analysis in science.
Inspired by the path of achievement of AI, we have developed a method to construct accurate ground state wave functions of multiquark configurations within a quark model. 
We successfully tested our method through comparisons with meson-type two-body systems with analytic and numerical solutions.
We then applied our method to find the ground-state solutions of $T_{cc}(ud \bar{c} \bar{c})$ and $T_{bb}(ud \bar{b}\bar{b})$ states.  Our findings indicate that our approach outperforms existing methods, achieving greater accuracy in reproducing highly intricate configurations.  Within the model parameters, we find that the $T_{cc}$ is a compact multiquark configuration.
\end{abstract}

\maketitle

\section{Introduction}

As science evolves, we have witnessed that it is becoming increasingly important not only in the industrial field but also in the scientific field to calculate exact values efficiently and reliably when only numerical solutions are available. 
Navie numerical methods are sometimes time-consuming and their reliability is questionable.  One area of physics where efficient but reliable numerical methods have become essential is the quark model.  Historically, quark models have primarily been applied to mesons (2-body) or baryons (3-quark) exclusively. However, with the recent discovery of many exotic hadrons~\cite{Chen:2022asf}, there is now a need to determine the exact ground state wave function of tetra-quark, pentaquark, and dibaryon systems to assess whether these multi-quark states are indeed compact quark states. 

Here, we develop a numerical algorithm, inspired by how artificial intelligence (AI) methods have evolved, to lay the ground for finding the exact numerical wave function for multiquark states.  

While the terminology 'AI' may be relatively recent, its origins can be traced back to basic data science in general physics experiments, particularly using techniques like least-square fitting (LSF). Similar to LSF, AI involves finding optimal parameter sets through appropriate structures (architectures) to capture trends in data. Just as LSF aided Hubble in discovering the accelerating universe nearly a century ago~\cite{Hubble:1929ig}, recent advancements in AI are shedding light on the atomic world, demonstrating a successive path of development. 

One can think of this trend as the optimization of numerous parameters within the appropriate architecture. The most challenging aspect lies in constructing the proper architecture, such as a convolutional neural network~\cite{6795724} in computer vision and a transformer~\cite{46201} in large language models. 
However, this process often entails a considerable amount of trial and error, without any assurance of a definitive solution. In physics, however, we frequently begin with an ansatz and then can leverage empirical observations from AI

The paper is organized as follows. In Sec.~\ref{formalism}, we introduce the method to build the most accurate wave function based on observations from AI. The effectiveness of our method is then tested by comparing the results with those of a Coulombic system and the $J/\psi$ in a non-relativistic quark model. Secondly, we describe the algorithm implementing our method in Sec.~\ref{algorithm}.  We then apply our algorithm to $T_{cc}$ and $T_{bb}$ tetraquark systems in Sec.~\ref{application}. Finally, we illustrate the procedure for constructing the ansatz for $T_{cc}$ and $T_{bb}$. Physical considerations are provided in Appendices A.

\section{Construction of the wave function}\label{formalism}

Our method draws inspiration from AI, where models of the same architecture trained on the same dataset exhibit various parameter values resulting in similar performance metrics, such as test accuracy in classification tasks. This observation suggests the presence of both flat and bumpy local extrema in the optimization process.  
Furthermore, OpenAI demonstrated its first profitable model by increasing the number of parameters by 
approximately 100-fold, from GPT-2 to GPT-3. This suggests that model performance improves with an increase in the number of parameters within the appropriate architecture.
Following these empirical observations, we have devised a method to construct the exact wave function by combining many parameterized approximate wave functions (Eq.~(\ref{eq:method})), typically referred to as an ansatz.
\begin{equation}\label{eq:method}
    \Psi = \sum_s \psi_{s\{s_i\}},
\end{equation}
where $s$ denotes different trial wave functions and $\{ s_i \}$ the different parameters in each wave function.
Exploiting this freedom, it is advantageous to choose an ansatz in harmony with the potentials in the consideration, which means the potentials have analytic expression when expectation values are taken. Otherwise, it is inevitable to conduct numerical integration that accompanies huge computation time as well as accumulative error.

\subsubsection{Coulomb potential}
Let us first demonstrate our method of obtaining the wave function by applying it to the Coulomb potential, for which the analytic solution is known.  
We study the following two-body Coulomb potential problem.
\begin{equation}
    H= \sum_{i = 1}^2\bigg( m_i + \frac{p_i^2}{2m_i} \bigg) - \frac{3}{4}\sum_{i<j}^2\frac{\lambda_i^c}{2}\frac{\lambda_j^c}{2}\Big( -\frac{\kappa}{r_{ij}} \Big).
\end{equation}
Here, we introduced the color matrix $\lambda^c_i$ for particle $i$ to facilitate a straightforward transition to QCD problems later. We chose the charm and light quark masses in the quark models: $m_1 = m_c = 1922\text{ MeV}$, $m_2 = m_q = 342.0\text{ MeV}$, and $\kappa = 120.0\text{ MeV} \cdot \text{fm}$. All parameters are the same as in Ref.~\cite{Noh:2021lqs}. The analytic solution yields a binding energy of $-53.686$ MeV and the ground state wave function as given below.
\begin{equation}\label{eq:Coulomb}
    R(r) = 2\bigg( \frac{1}{a_0'} \bigg)^{\frac{3}{2}}e^{-\frac{r}{a_0'}}.
\end{equation}
Here, $r_ = |\vec{r_1} - \vec{r_2}|$ and $\mu = \frac{m_1m_2}{m_1 + m_2}$, 
and $a_0' = \frac{\hbar}{\mu c \alpha'} = \frac{\hbar^2}{\mu \kappa}$ where $\alpha' = \frac{\kappa}{\hbar c}$. Now, if one did not know about the analytic solution of the ground state of the Coulomb potential, one might use a parameterized Gaussian ($e^{-sr^2}$) as an ansatz based on the nodal theorem and anticipate approaching the exact solution by adding it infinitely.  

\begin{figure}[h]
\centering
\includegraphics[width=0.9\linewidth]{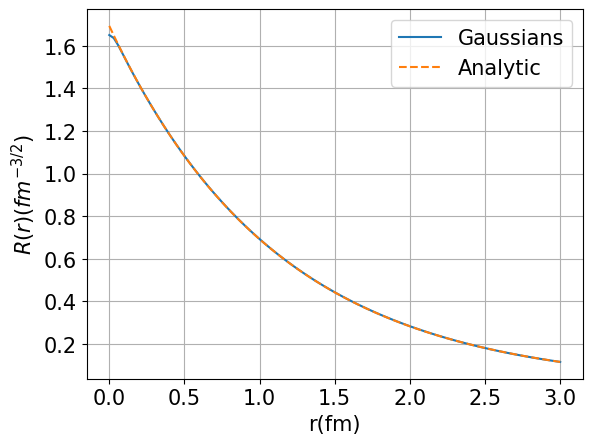}
\caption{Comparison between analytic and constructed wave functions with 97 Gaussians.}
\label{far_exact}
\end{figure}

\begin{figure}[h]
\centering
\includegraphics[width=0.9\linewidth]{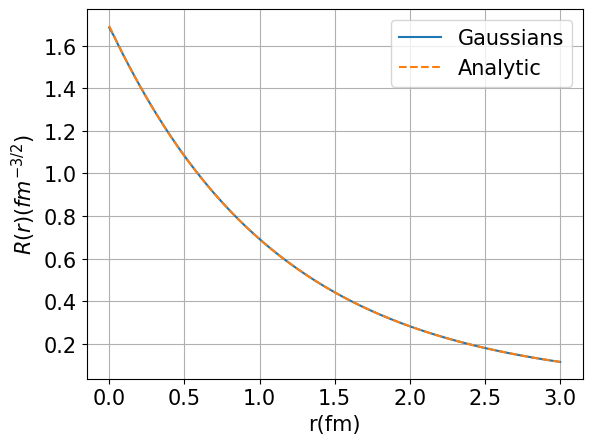}
\caption{Comparison between analytic and constructed wave functions with 427 Gaussians.}
\label{near_exact}
\end{figure}

Comparisons between the wave functions obtained by the method and the analytic wave function are shown in Fig.~\ref{far_exact} when 97 Gaussians are used, and in  Fig.~\ref{near_exact}, when 427 Gaussians are used.  The binding energies are given as $-53.663$ MeV and $-53.684$ MeV, respectively.

One notes that the tiny discrepancy between the analytic and constructed wave functions in Fig.~\ref{far_exact} near the origin can no longer be discerned in Fig.~\ref{near_exact}. Quantitatively, the relative error is reduced by 91\% as one increases the Gaussians from 97 to 427. 

Therefore, we have confirmed that with a sufficient number of Gaussians, the method can construct an accurate wave function comparable to the analytic one, thereby correctly characterizing the physical system.

\subsubsection{Charmonium states}

We now test our method to solve for the mass of the $J/\Psi$ in a quark model, for a case with only a numerical solution. The Hamiltonian and parameters are the same as in Ref.~\cite{Noh:2023zoq}.
\begin{equation}
    H= \sum_{i = 1}^2\bigg( m_i + \frac{p_i^2}{2m_i} \bigg) - \frac{3}{4}\sum_{i<j}^2\frac{\lambda_i^c}{2}\frac{\lambda_j^c}{2}\Big(V_{ij}^C + V_{ij}^{CS} \Big),
\end{equation}
where
\begin{equation}
    V_{ij}^C = -\frac{\kappa}{r_{ij}} + \frac{r_{ij}}{a_0^2} - D,
\end{equation}
\begin{equation}
    V_{ij}^{CS} = \frac{\hbar^2c^2\kappa '}{m_im_jc^4}\frac{e^{-r_{ij}/r_{0ij}}}{(r_{0ij})r_{ij}}\sigma_i\cdot\sigma_j,
\end{equation}
\begin{equation}
    r_{0ij} = \frac{1}{\alpha + \beta\frac{m_im_j}{m_i + m_j}},\quad \kappa' = \kappa_0\bigg( 1 + \gamma\frac{m_im_j}{m_i + m_j} \bigg).
\end{equation}
Here $m_c=1895$ MeV, $\kappa = 97.7$ MeV$\cdot$fm, $a_0 = 0.0327338 \text{ (MeV}^{-1}\text{ fm})^{1/2}\text{, } D = 959\text{ MeV, } \alpha = 1.1349\text{ fm}^{-1}\text{, } \beta = 0.0011554\text{ (MeV fm)}^{-1}\text{, } \gamma = 0.001370\text{ MeV}^{-1}\text{, and }\kappa_0 = 213.244\text{ MeV}$.

\begin{figure}[h]
\centering
\includegraphics[width=0.9\linewidth]{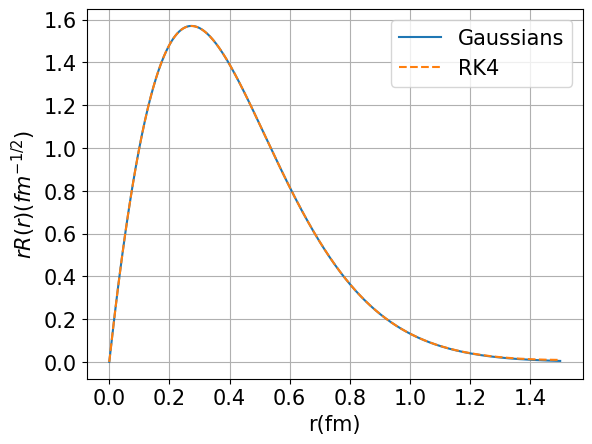}
\caption{Comparison between wave functions using RK4 and this method.   }
\label{mine}
\end{figure}

Fig.~\ref{mine} shows the comparison of the wave function obtained by 4th order numerical solution(RK4) to that obtained with 70 Gaussians  ($e^{-sr^2}$).  As can be seen in the figure, the obtained wave functions are indistinguishable.  The binding energy used in the RK4 method is E=$-660.464$ MeV, which coincides with the one obtained by our method.

\section{Optimization Algorithm}\label{algorithm}

In Fig.~\ref{near_exact}, we employed 400 Gaussians, each with one parameter. Conducting a search without any strategy would require $n^{400}$ calculations of the eigenvalue, with n search points per parameter. Hence, the introduction of an optimization algorithm into our method is necessary.

The choice of algorithm will depend on the specific problem being analyzed. In the context of the quark model, we observe the following. 
Firstly, the system can be effectively described as a varying function with respect to the parameters in the wave function. For example, when calculating the ground state of $T_{cc}$, the unitary matrix that diagonalizes the Hamiltonian can only be obtained after determining each matrix element of the Hamiltonian with a chosen set of parameters.  Consequently, if one applies the gradient descent method, it would operate with the previously determined unitary matrix, resulting in a stall of the calculation away from the ground state.
Secondly, during such parameter searches, when additional parameters are introduced to enhance accuracy, the values of the existing set undergo minimal changes relative to their absolute values. Finally, the number of parameters required for the calculated physical values to converge is in the order of hundreds, in contrast to the millions needed, for example, in systems related to neural networks

Hence, the first observation suggests that the algorithm should be function-agnostic.
Secondly, accurate determination of physical parameters can be achieved through small adjustments to the optimized parameters. Finally, it is legitimate to iteratively search nearby points in the selected sub-parameter space from the previously optimized point. Below, in Algorithm 1, we provide a general description of the algorithm that encompasses the above criteria, following the spirit of Occam's razor.\\

\begin{algorithm}[H]
\caption{Parameter Optimization Algorithm for Physical Systems}
\label{alg:param_opt}
\begin{algorithmic}[1]
\State Initialize parameters $P = \emptyset$
\Repeat
    \State Add new parameters $P^{\text{new}}$ to $P$
    \State Brute search on $P^{\text{new}}$
    \State $P^{\text{new}} \gets$ The most optimized $P^{\text{new}}$
    \Repeat
        \State Perturb $P^{\text{new}}$ by changing some fractions times -1, \\ \quad \quad \quad 0, and 1
        \State $P^{\text{new}} \gets$ The most optimized $P^{\text{new}}$
    \Until{Convergence}
    \Repeat
        \State Randomly select parameters $P^{\text{rand}}$ from $P$
        \State Perturb $P^{\text{rand}}$ by changing some fractions times -1, \\ \quad \quad \quad 0, and 1
        \State $P^{\text{rand}} \gets$ The most optimized $P^{\text{rand}}$
    \Until{Convergence}
\Until{Convergence}
\end{algorithmic}
\end{algorithm}

In addition to its simple and intuitive implementation, the 'accuracy' is statistically confirmed through the usual derivative sense, as demonstrated in the application to the tetraquark system in the next section.

\section{Application to $T_{cc}$ and $T_{bb}$}\label{application}

Recently, the LHC observed a flavor-exotic state $T_{cc}$, composed of two charm quarks and two anti-up and anti-down quarks ($cc \bar{u}\bar{d}$)~\cite{LHCb:2021vvq}. This state was predicted long ago in the quark model based on strong diquark attraction~\cite{Ballot:1983iv, Zouzou:1986qh}. However, it could also potentially be a molecular structure composed of weakly bound $DD^*$ particles~\cite{Tornqvist:1993ng, Tornqvist:2004qy}. The ambiguity arises because the observed state lies very close to the mesonic threshold. Quark model calculations yield results both below and above the threshold, depending on slight changes in the model parameters and the calculation method employed to determine the wave function.
The expected quantum number we assume for $T_{cc}$ is $I(J^P)=0(1^+)$.

Here, the goal of the problem is to find the minimum ground state energies of tetraquark systems $T_{cc}$ and $T_{bb}$ under the condition that the total quark wave function obeys Pauli exclusion principle. 
Since the $(\bar{u}\bar{d})$ quarks are in the isospin=0 state, the flavor part of the wave function for the light quarks should be antisymmetric under the exchange of two particles. The flavor part of the two $(cc)$ should be symmetric.  Then the necessary symmetry property under the exchange to satisfy the Pauli Principle for the possible configurations are given in Appendix A. The corresponding spatial wave functions that satisfy certain symmetry properties are also given there.  
With these ansatzes, one can separate the algorithm into three steps as detailed in the following subsections.

\subsection{Add new parameters and brute search}

To ensure the wave function is consistent with physics at all stages, it is necessary to introduce six parameters ($C_{11}, C_{22}, C_{33}, C_{12}, C_{13}, \text{ and } C_{23}$) with certain signs into the four Gaussians. This is done to satisfy the Pauli principle, as defined in Appendix A. The diagonal parameters($C_{11}, C_{22}, \text{ and } C_{33}$) multiply the square of internal coordinates in the Gaussian and could become large when the parameter space is not bumpy. The off-diagonal parameters($C_{12}, C_{13}, \text{ and }C_{23}$) multiply products of internal coordinates and represent effects from internal angular dependence of internal coordinates\cite{Park:2023ygm}.  
Therefore, there are exponentially spaced grids for the diagonal parameters while the grids for the off diaonal parameters are equally spaced as given in subalgorithm~A.

\begin{algorithm}[H]
  \caption*{Subalgorithm~A}
  \label{alg:brute_force}
  \begin{algorithmic}[1]
    \State{Add $P^{new} = \{C_{11}, C_{22}, C_{33}, C_{12}, C_{13}, C_{23}\}$ to $P$}
    \For{$i_{11},~i_{22},~i_{33},~i_{12},~i_{13},~i_{23} = 1$ to $6$}
        \State $C_{11},~C_{22},~C_{33} = 2^{i_{11}},~2^{i_{22}},~2^{i_{33}}$
        \State $C_{12},~C_{13},~C_{23} = -3.5 + i_{12},~-3.5 + i_{13},~-3.5 + i_{23}$  
        \State $P^{\text{new}} \gets$ The most optimized $P^{\text{new}}$
    \EndFor
  \end{algorithmic}
\end{algorithm}

\subsection{Perturb the new set}

After finding a new parameter set in the grid, the next step is to perturb that set. This procedure is accomplished by implementing Subalgorithm B, with the searching fraction set to $\frac{1}{5^{cnt}}$, where $cnt$ ranges from 2 to 4. In other words, it searches for the most optimized point in the 6-dimensional parameter space by adjusting $\frac{1}{5^{cnt}}$ of the parameters from the previous most optimized point. The tolerance for no update is set to 1, meaning that if $\frac{1}{5^{2}}$ yields the same energy as before, the fraction becomes $\frac{1}{5^{3}}$.
It is shown that $\frac{\partial E}{\partial \ln P^{\text{new}}}$ = 0 is confirmed by the presence of 0 for $j$ at the fourth line of subalgorithm~B.
\begin{algorithm}[H]
  \caption*{Subalgorithm~B}
  \label{alg:param_search}
  \begin{algorithmic}[1]
  \State $cnt = 2$
    \While{$cnt \leq 4$}
      \For{$i = 1 \text{ to } 6$}
        \For{$j = -1, 0, 1$}
          \State Search with $P^{\text{new}}_i \mathrel{+}= j \times P^{\text{new}}_i/5^{cnt}$
        \EndFor
      \EndFor
          \State $P^{\text{new}} \gets$ The most optimized $P^{\text{new}}$
      \If{no change in $P^{\text{new}}$}
        \State $cnt\mathrel{+}=1$
      \EndIf
    \EndWhile
  \end{algorithmic}
\end{algorithm}

\subsection{Perturb the whole set}

\begin{algorithm}[H]
  \caption*{Subalgorithm~C}
  \label{alg:param_search_all}
  \begin{algorithmic}[1]
  \State $tol_{cnt}$ = 0, $update = 1$
    \While{$update$}
    \State $cnt = 2$, $update = 0$
    \While{$cnt \leq 5$}
    \State Pick $P^{\text{rand}} = \{P^{\text{rand}}_i\}$ from $P$
      \For{$i = 1 \text{ to } \lVert P^{\text{rand}} \rVert$}
        \For{$j = -1, 0, 1$}
          \State Search with $P^{\text{rand}}_i \mathrel{+}= j \times P^{\text{rand}}_i/5^{cnt}$
        \EndFor
      \EndFor
          \State $P^{\text{rand}} \gets$ The most optimized $P^{\text{rand}}$
      \If{no change in $P^{\text{rand}}$}
        \State $tol_{cnt}\mathrel{+}=1$
        \If{$tol_{cnt} > f(cnt, \lVert P \rVert)$}
          \State $cnt\mathrel{+}=1$
          \State $tol_{cnt} = 0$
        \EndIf
      \Else 
        \State $update = 1$
      \EndIf
    \EndWhile
    \EndWhile
  \end{algorithmic}
\end{algorithm}

In contrast to subalgorithm B, a tolerance ($tol_{cnt}$) exists for each $cnt$ given by a function of $cnt$ and the number of parameters in subalgorithm C due to stochasticity. Furthermore, even if the parameters change only once in all $cnt$, the whole process is repeated to statistically assure $\frac{\partial E}{\partial \ln P} = 0$ checked by the variable $update$. $update$ becomes 1 from 0 when the parameters are more optimized, so there must be no update of the parameters to terminate subalgorithm~C.

By iterating subalgorithms~A,~B, and~C until convergence is reached, one can build the precise wave function resulting in the correct physical values within the hypothesized model. The results of our calculation using a total of 150 parameters are given in Table.~\ref{tetra_results} and compared with those obtained using semi-complete Harmonic oscillator bases given in Ref.~\cite{Noh:2023zoq}. The Hamiltonian and parameters used in Ref.~\cite{Noh:2023zoq} are the same as given in Eqs.~(4)$\sim$(7). 
It is worth noting that our result yields a mass for $T_{cc}$ approximately 0.3\% smaller than that reported in Ref.~\cite{Noh:2023zoq}. Furthermore, the relative distances are found to be about 10\% larger.

\begin{table}[h]
\centering
\begin{tabular}{|c|c|c|c|c|r|}\hline
& Tetraquark & Mass(MeV) & $r_{12}$(fm) & $r_{23}$(fm) & $r_{34}$(fm)\\ \hline
Method& $T_{cc}$& 3861 & 0.98 & 0.76 & 0.77 \\ \cline{2-6}
& $T_{bb}$ & 10461 & 0.71& 0.62 & 0.29\\ \hline
Ref.~\cite{Noh:2023zoq} & $T_{cc}$ & 3872 & 0.87 & 0.71 & 0.64 \\ \cline{2-6}
& $T_{bb}$ & 10464 & Non & Non & Non\\ \hline
\end{tabular}
\caption{$1,2$ denotes the two light quarks and $3,4$ are the two heavy antiquarks. $r_{ij}$ deonotes the relative distance between particle $i$ and $j$. $r_{13}$, $r_{14}$, and $r_{24}$ are the same as $r_{23}$ because the wave function obey Pauli principle.}
\label{tetra_results}
\end{table}

\begin{figure}[h]
\centering
\includegraphics[width=0.9\linewidth]{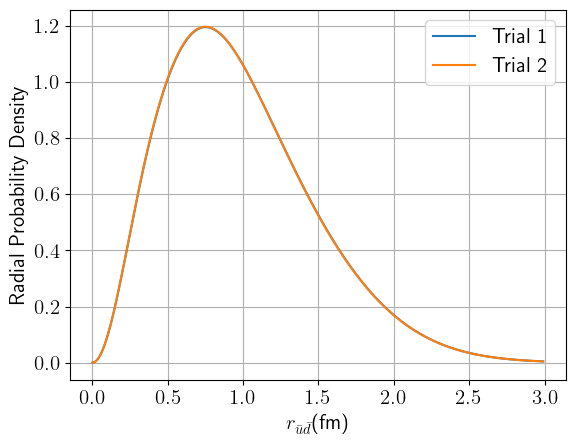}
\caption{Radial probability densities between $\bar{u}$ and $\bar{d}$ for $T_{cc}$. Both trials have the same number of parameters (150) but different values.}
\label{ud_P}
\end{figure}

\begin{figure}[h]
\centering
\includegraphics[width=0.9\linewidth]{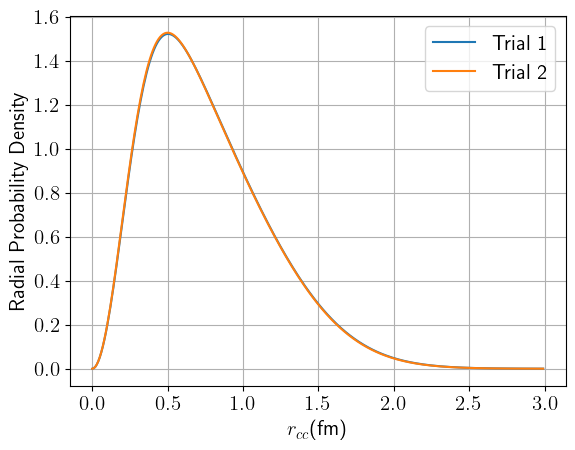}
\caption{Radial probability densities between $c$ and $c$ for $T_{cc}$. Both trials have the same number of parameters (150) but different values.}
\label{cc_P}
\end{figure}

As inferred in Sec.~\ref{formalism}, our method yields almost identical physical values despite using different parameter sets, as long as each calculation goes through the algorithm with the same number of parameters. 
The second and thrid line in Table~\ref{trials} show calculations for the $T_{cc}$ with the same number of parameters but with slightly different parameters. Despite noticeable differences in the values of the parameters (by 16\%) the physical values remain almost the same.  The wave functions also remain almost identical as can be seen in Fig.~\ref{ud_P} and ~\ref{cc_P}.  
The fourth line of Table~\ref{trials} shows the saturated value after using more than 200 parameters. The calculation with 216 parameters yields $3858.51$ MeV of the ground state mass for $T_{cc}$, which is below the calculated $D,D*$ threshold of 3874 MeV obtained with the same Hamiltonian(Eqs.~(4)$\sim$(7)).

\begin{table}[h]
\centering
\begin{tabular}{|c|c|c|c|r|}\hline
\# of parameters & Mass(MeV) & $r_{12}$(fm) & $r_{23}$(fm) & $r_{34}$(fm)\\ \hline
 150& 3861.47 & 0.985& 0.756 & 0.768 \\ \hline
 150 & 3861.56 & 0.982& 0.755 & 0.765\\ \hline
 216 & 3858.51 & 0.992& 0.761 & 0.774\\ \hline
\end{tabular}
\caption{Calculations for $T_{cc}$ with different parameter sets (lines 2 and 3) and different number of parameters used (line 4). }
\label{trials}
\end{table}

\section{Conclusion}
 
Inspired by the evolution of artificial intelligence, we have developed a method to construct the ground state wave functions in a quark model. We have tested the convergence of the wave function to the analytic solutions. Furthermore, we have used the method to accurately estimate the masses of the recently observed 
$T_{cc}$ and $T_{bb}$ sates.

We have emphasized the importance of accurately determining the wave function in multiquark systems. A poorly determined wave function can yield a similar mass value but result in different sizes, leading to varying predictions about the nature of exotic hadron states. There are many newly observed exotic hadrons \cite{Chen:2022asf}, and determining the nature of these exotic states is the next step we need to take. Therefore, it is undeniable that this method is essential for contemporary science.

Notable features of our method are as follows. First, there is no accumulative error in contrast to iterative solvers such as Rugge-Kutta method. Our method improves the accuracy of the wave function by iteratively adding parameters based on the variational principle. Secondly, the method does not require heavy numerical coding. 
Once the correct structure of the wave function is given, the parameters begin to adjust dynamically to reach the ground state. Finally, it has high extensibility. For example, even if potentials were given numerically, one could fit these potentials with parameterized known functions which have analytic expressions when expectation values are taken by the assumed wave function. Then, as usual, one can see what physical values these numerical potentials predict after going through the algorithm.

However, there is a room for improvement in our method. The algorithm employed is not the best one. Even though it is working well, it still involves brute-search in the sub-parameter space. In that variants of the gradient descent have been actively investigated~\cite{pmlr-v28-sutskever13, adam}, incorporating them in the algorithm might drastically improve the convergence rate.

Despite the room for improvement, our method can now be used to study more complex systems such as pentaquark dibaryons, which will involve many more quark states and hence an extended form of the wave function.

\appendix

\section{Ansatz for $T_{cc}$ and $T_{bb}$ ground states}
When the quantum numbers of $T_{cc}$ and $T_{bb}$ are assumed to be $I(J^P)=0(1^+)$\cite{Vijande:2003ki}, allowing for all possible states satisfying the Pauli principle, the total number of bases is 6. Since there are definite sign changes after permutations of constituent quarks(or antiquarks) for the color, spin, and flavour wave functions~\cite{Park:2013fda}, one can determine the sign change for the spacial wave function to satisfy the Pauli principle.  These sign changes are shown in Table.~\ref{signs_permu}.

\begin{table}[h]
\centering
\begin{tabular}{|c|c|c|c|c|c|c|}\hline

color & $\bar{3}3$ & $\bar{3}3$ & $\bar{3}3$ & $6\bar{6}$ & $6\bar{6}$ & $6\bar{6}$ \\ \hline
spin & 10 & 11 & 01 & 10 & 11 & 01 \\ \hline
isospin & 00 & 00 & 00 & 00 & 00 & 00 \\ \hline
signs without space& $++$ & $+-$ & $--$ & $--$ & $-+$ & $++$\\\hline
required signs for space & $--$ & $-+$ & $++$ & $++$ & $+-$ & $--$\\\hline
total signs& $--$ & $--$ & $--$ & $--$ & $--$ & $--$\\\hline

\end{tabular}
\caption{States and signs after applying permutations in the inner-product bases. The first state and sign in each item refer to the case of light quarks and those of the second to heavy antiquarks.}
\label{signs_permu}
\end{table}

We take the ansatz consistent with Pauli principle. For that purpose, we introduce the matrix and internal coordinates between constituent quarks, as shown in Eq.~\ref{structure}. $\vec{r_1}$ and $\vec{r_2}$ denote quarks and $\vec{r_3}$ and $\vec{r_4}$ do heavy antiquarks in Eq.~\ref{inter_coordi} for internal coordinates.  Using these, the Gaussian wave function can be simply expressed as $\exp(-X^TC^sX)$.

\begin{equation}
   C^{s} = \begin{pmatrix} 
   C^{s}_{11} & C^{s}_{12} & C^{s}_{13}  \\
   C^{s}_{12} & C^{s}_{22} & C^{s}_{23}  \\
   C^{s}_{13} & C^{s}_{23} & C^{s}_{33}  \\
   \end{pmatrix}, \quad
X = \begin{pmatrix}
\vec{\rho}\\
\vec{\rho'}\\
\vec{x}\\
\end{pmatrix}
\label{structure}
\end{equation}

\begin{equation}
\vec{\rho} = \frac{\vec{r_1} - \vec{r_2}}{\sqrt{2}},\quad \vec{\rho'} = \frac{\vec{r_3} - \vec{r_4}}{\sqrt{2}}, \quad \vec{x} = \frac{\vec{r_1} + \vec{r_2} - \vec{r_3} - \vec{r_4}}{2}
\label{inter_coordi}
\end{equation}

Therefore, the signs of internal coordinates change after permutations between quarks and heavy antiquarks as given in Eq.~\ref{coordi_permu}. To recover the appropriate symmetries, the sign changes are transferred to the parameters (Eqs.~\ref{12_G}$\sim$\ref{12_34_P}).

\begin{eqnarray}
(12)X = \begin{pmatrix}
-\vec{\rho}\\
\vec{\rho'}\\
\vec{x}\\
\end{pmatrix}, ~~
(34)X = \begin{pmatrix}
\vec{\rho}\\
-\vec{\rho'}\\
\vec{x}\\
\end{pmatrix}, \nonumber \\
(12)(34)X = \begin{pmatrix}
-\vec{\rho}\\
-\vec{\rho'}\\
\vec{x}\\
\end{pmatrix}.
\label{coordi_permu}
\end{eqnarray}

\begin{equation}
(12)\exp(-X^TC^{s}X) = \exp(-X^TC^{s}_{(12)}X)
   \label{12_G},
\end{equation}
where
\begin{equation}
C^{s}_{(12)} = \begin{pmatrix} 
   C^{s}_{11} & -C^{s}_{12} & -C^{s}_{13}  \\
   -C^{s}_{12} & C^{s}_{22} & C^{s}_{23}  \\
   -C^{s}_{13} & C^{s}_{23} & C^{s}_{33}  \\
   \end{pmatrix}.
   \label{12_P}
\end{equation}

\begin{equation}
(34)\exp(-X^TC^{s}X) = \exp(-X^TC^{s}_{(34)}X),
   \label{34_G}
\end{equation}
where
\begin{equation}
C^{s}_{(34)} = \begin{pmatrix} 
   C^{s}_{11} & -C^{s}_{12} & C^{s}_{13}  \\
   -C^{s}_{12} & C^{s}_{22} & -C^{s}_{23}  \\
   C^{s}_{13} & -C^{s}_{23} & C^{s}_{33}  \\
   \end{pmatrix}.
   \label{34_P}
\end{equation}

\begin{equation}
(12)(34)\exp(-X^TC^{s}X) = \exp(-X^TC^{s}_{(12)(34)}X)
   \label{12_34_G},
\end{equation}
where
\begin{equation}
C^{s}_{(12)(34)} = \begin{pmatrix} 
   C^{s}_{11} & C^{s}_{12} & -C^{s}_{13}  \\
   C^{s}_{12} & C^{s}_{22} & -C^{s}_{23}  \\
   -C^{s}_{13} & -C^{s}_{23} & C^{s}_{33}  \\
   \end{pmatrix}.
   \label{12_34_P}
\end{equation}

As a result, the required spatial wave function for certain permutations for both the light quarks and heavy antiquarks are given as follows.

\begin{itemize}

\item{\bf For $++$ permutation symmetry:}

$\psi_{s\{s_i\}} = \exp(-X^TC^{s}X) + \exp(-X^TC^{s}_{(12)}X) + \exp(-X^TC^{s}_{(34)}X) + \exp(-X^TC^{s}_{(12)(34)}X)$.

\item{\bf For $+-$ permutation symmetry:}

$\psi_{s\{s_i\}} = \exp(-X^TC^{s}X) + \exp(-X^TC^{s}_{(12)}X) - \exp(-X^TC^{s}_{(34)}X) - \exp(-X^TC^{s}_{(12)(34)}X)$.

\item{\bf For $-+$ permutation symmetry:}

$\psi_{s\{s_i\}} = \exp(-X^TC^{s}X) - \exp(-X^TC^{s}_{(12)}X) + \exp(-X^TC^{s}_{(34)}X) - \exp(-X^TC^{s}_{(12)(34)}X)$.

\item{\bf For $--$ permutation symmetry:}

$\psi_{s\{s_i\}} = \exp(-X^TC^{s}X) - \exp(-X^TC^{s}_{(12)}X) - \exp(-X^TC^{s}_{(34)}X) + \exp(-X^TC^{s}_{(12)(34)}X)$.

\end{itemize}

\section*{ACKNOWLEDGEMENTS}

This work was supported by the Korea National Research Foundation under Grant Nos. 2023R1A2C300302311 and 2023K2A9A1A0609492411.

\bibliography{main.bib}
\end{document}